\documentclass[12pt,authoryear]{elsarticle}

\usepackage{siunitx}
\usepackage{lineno,hyperref}
\modulolinenumbers[5] 
\usepackage{eurosym}
\usepackage{amsmath}
\usepackage{mathbbol}
\usepackage{enumitem}
\usepackage{graphicx}
\usepackage{epsfig}
\usepackage{colortbl} 
\usepackage{xcolor} 
\usepackage{booktabs}
\usepackage{epstopdf}
\usepackage{amssymb,latexsym,bm}
\usepackage{natbib}
\usepackage{hyperref}
\usepackage{amsfonts}
\usepackage{dcolumn}
\usepackage{subfig}
\usepackage{arydshln}
\usepackage{color,soul}
\usepackage{mathtools}
\usepackage{multirow,adjustbox}
\usepackage{floatrow}
\hypersetup{ colorlinks = true,  linkcolor = blue, anchorcolor = blue, citecolor = blue, filecolor = blue, urlcolor = blue}

\usepackage{hyphenat}
\definecolor{blue1}{RGB}{84,3,30}

\makeatletter
\def\ps@pprintTitle{%
  \let\@oddcultural\let\@evenhead\@empty
  \let\@oddfoot\@empty
  \let\@evenfoot\@oddfoot
}
\makeatother

\begin{document}


\begin{frontmatter}

\title{\textbf{Gender politics, environmental behaviours, and local territories:  Evidence from Italian municipalities}}

\author[inst1]{Chiara Lodi}

\author[inst1]{Agnese Sacchi}

\author[inst1]{Francesco Vidoli}
\address[inst1]{University of Urbino Carlo Bo, Department of Economics, Society, Politics, Via Aurelio Saffi, 42, Urbino 61029, PU Italy}

\begin{abstract}
We investigated the impact of female politicians on waste collection in Italian municipalities in different territories observed over the years 2010-2019. We used a staggered difference-in-differences design to obtain a causal interpretation of the estimated effects. We find that the majority of women in the municipal council positively influence pro-environmental individual behaviour. The impact of a female-majority council is heterogeneous by region and more pronounced in areas with lower social capital. Female politicians as catalysts for positive change fade after 5-6 years, likely due to persistent social norms locally, thus stressing the need for additional cultural actions with long-lasting effects.
\end{abstract}



\end{frontmatter}







\section{Introduction}
\label{intro}

Where do policies, public interventions, and historical human events stand? we believe that they are rooted in the cultural foundation of a territory, enabling them to evolve over time and produce outcomes if the conditions are conducive; for which reason, a territory might react to changes, possibly with intensity, yet this reaction may prove unproductive. Social norms, civil capital, culture, and institutional setting could be crucial in this framework \citep{guiso_etal_2006, guiso_etal_2011, acemoglu2021}. Therefore, to assess the impact of a cultural change on local territories, it could be helpful to interpret the territory as a set of social conventions rather than a simple administrative unit; this means moving away from the idea that a geographic area is determined ex ante by normative criteria only. As a result, assessing the impacts of various policies on a territory necessitates acknowledging that norms and conventions may vary across regions, and thus, they can be represented not only by artificial geographies\footnote{We use "artificial" term as originally presented by Hobbes, every State, region, province, or administrative unit could precisely be defined as an artificial construct, such as a political machine built by the agreement of men \citep{ellis1971}. At the same time, family and its social conventions would represent the natural state of men.} but also through other dimensions.

Based on these arguments, we assess the effects of women in politics, as a culturally relevant change in recent years, on public place-based policies, such as waste recycling management, at the municipal level in Italy over the years 2010-2019. Specifically, we first estimate the impact of having a female mayor or a municipal council mainly composed of women on the waste recycling rate and then test if this impact is similar in different territories, i.e., spatial stationary.

In recent years, the study of the positive effect of female policy makers on environmental matters has increased. It may be influenced by multiple variables, such as women's roles, lifestyle, and the social norms to which they adhere \citep{wut2021, wilde2022, desario2023, stef2023}. However, further investigation is needed to understand whether these positive pro-environmental behaviours depend on gender or if they only reflect confounding variables. In addition, there may be a selection bias, as mayors who are less effective in waste management tend to govern municipalities where residents are less vigilant about environmental protection. To obtain unbiased estimates and a causal interpretation of the estimated effect, we use the staggered difference-in-differences design (CSDID) proposed by \citealp{CALLAWAY2021200}. This empirical approach allows for a comparison of municipalities that have the same political, environmental, and gender preferences while taking into account the fact that the elections were held at different times. 

We rely on Italian municipalities that serve as a laboratory example for several reasons. First, they perform a wide range of functions, including education and social services, the environment, and the territory, thus providing many local public services (\emph{e.g.}, school buses, street lighting, waste collection, etc.), affecting the quality of life and well-being of citizens \citep{ermini2023}. In particular, local investments in territorial development, green transition, and environmental protection have increased over time and have recently been further stimulated by the Next Generation EU programme \citep{rivas2021, ec2023}. In this framework, the mayor and the municipal council manage those specific functions/policies pertaining to its area and accountable for the administration of the municipality. Second, gender politics and related institutional reforms have stimulated interest in investigating the impact of female politicians on public spending and other political outcomes, especially at the local level \citep{depaola2010, rigon2012, baltrunaite2014, braga2017, baltrunaite2019, cella2023}. 

Our results offer new insights into the existing literature. First, we provide causal evidence for the effect of female representation on citizen green behaviour. This occurs not mainly when a female mayor is elected, but when the majority of the local council comprises women. In other words, a single person cannot trigger a strong change in its recycling behaviours that instead appears to occur when a critical mass is at work. Second, such a change is statistically significant but tends to fade in the medium term; this may be due to our limited time frame, and, more likely, more structural behaviour seems to return in the medium term. Third, the effect of recycling rate change due to a female-majority council is not stationary in space, but is stronger in regions with lower social capital. Our evidence supports previous findings linking behaviour, civic sense, and social capital in a macro-perspective depending on the local territory \citep{Putnam1993, Fukuyama1995, calcagnini2019, acemoglu2021}. We highlight that every change is shown to lie in a specific territory. Therefore, it is necessary to focus on \textit{where} and not only \textit{how} such change takes place without relying on predetermined administrative boundaries, which could not effectively capture the cultural basis. 

From a policy point of view, our results support the idea that gender roles shape attitudes and behaviour toward environmental and climate issues \citep{eige2023}, contributing to the literature on behavioural economics and social psychology on how gender is one factor that might influence pro-environmental behaviours \citep{raty2010, coelho2017, liu2019, bush2023}. As a corollary, alleviating the gender gap in politics and the power domain is also crucial to achieving the green transition goals, as the impact of climate and environmental change is not gender neutral \citep{ec2020, oecd2021}. 

The remainder of the article is structured as follows. Section \ref{design} briefly explains the research design and the mechanisms. Section \ref{back} describes the institutional background of Italian municipalities with respect to electoral rules and gender political reforms, then the regulations that ruled waste management and its separate collection. Sections \ref{method} and \ref{empirical} illustrate the empirical analysis, including the description of the data, the econometric framework, the main findings, and robustness checks. Section \ref{final} concludes, providing some policy recommendations.


\section{Research design and mechanisms}
\label{design}

In the modern economy, it is common to measure the effects of a policy in a detached manner from the social rules in which it is supported and in a territory often not adequately known; the latter would serve only as a passive backdrop to empirical analyses. In our study, the issue of \textit{where} should be understood for statistical sophistication but also as the core of the policy evaluation, originating its significance from the inner heterogeneity of individuals and their social habits. 

In Italy, there is evidence of different civic culture, social capital, trust, and ethical attitudes related to unequal participation in public, civil, and political life when comparing, for example, the southern and northern regions \citep{Putnam1993}. The persistence of such profound differences has also had inevitable implications for economics and politics \citep{calcagnini2019, mareetal2024}, making it difficult to depict these aspects with numerical data and imposing ex-ante geographies.

More importantly, a change in behaviours or cultural/social customs will likely have a locally differentiated effect, not necessarily uniform across territories. Gender norms might affect women and men's participation in the political arena \citep{dassonneville2021}, and the geographical context might also affect the effects of gender, if any, on environmental concerns in our case. Generally speaking, the observed gender gap in political participation could reflect the lower interest of women in politics, possibly arising from cultural perceptions of traditional gender roles and stereotypes, where politics is viewed as a male domain \citep{oecd2023}. Ultimately, this circumstance could have a different strength between territories and policy domains.

In terms of mechanisms, there is evidence that women politicians implement policies, more often than their male counterparts, in those areas of intervention that are traditionally seen as women's issues, such as childcare, health, social services and the environment \citep{clots2011, funk2015}. So far, much of the literature on the relationship between gender and environmental issues has focused on the private sector using micro-data by investigating the impact of women in executive or leadership roles on environmental policies and performance and the adoption of eco-innovations or other green practices at the firm level \citep{liu2018, birindelli2019, moreno2022}.

Gender disparities in attitudes about climate change and mitigation actions have recently been documented \citep{bush2023}. Evidence suggests that women show, on average, a greater awareness of environmental issues than men, leading the former to adopt behaviours aimed at protecting and preserving the environment as a global public good \citep{barsky1997, coelho2017, liu2019, desario2023}. For example, women are more likely to move towards more sustainable and "zero waste" household practices, in terms of energy consumption (cooking, cleaning) and recycling \citep{raty2010, wilde2022}, and participate in less visible environmental conservation efforts in the private sphere than men who are instead more likely to participate in more visible green behaviours \citep{thaller2020}. Furthermore, recent research in the European context highlights that girls and young women have emerged prominently as activists and leaders in climate movements \citep{noth2022}, calling for immediate action by those in decision-making positions as recognised in the EU Gender Equality Strategy 2020-2025.

Regarding Italy, recent evidence from the \cite{eige2023} shows that women are more inclined than men to choose environmentally friendly options, including feeling responsible for reducing climate change and avoiding plastic and/or single-use products. On the other hand, it should be noted that the ongoing green transition is expected to increase the demand for STEM topics and skills, where women are still underrepresented, making them not necessarily more sensitive to environmental issues.\footnote{Where the acronym STEM stands for \textit{Science, Technology, Engineering, and Mathematics}.} 

Given this framework, we shed new light on the relationship between gender in politics and citizens' environmental behaviours in heterogeneous geographic areas. In particular, our aim is to assess whether women's greater concern for environmental challenges translates into more political engagement and green actions from citizens, which can  increase separate waste collection in Italian municipalities in different territories. We also contribute to previous literature highlighting the impact of cultural values and social factors on pro-environmental attitudes, including waste management and recycling \citep{crociata2015, corraliza2000, knickmeyer2020}, in line with the "civic environmentalism" approach \citep{john1994civic, agyeman2003}.






\section{The institutional background}
\label{back}

In this section, we describe the institutional background of Italian municipalities regarding major and council elections and provide some information on separate waste collection at the municipal level.

\subsection{Electoral rules and gender politics in Italian municipalities}
\label{backComuni}

The municipal government is composed of the mayor and the executive committee (\textit{Giunta comunale}), appointed by the mayor and whose number of members depends on the municipality's population size.\footnote {The number ranges from 2 for municipalities with fewer than 3,000 inhabitants to 12 for those with over 1 million inhabitants, according to Law 56/2014.} The mayor and the executive body carry out the electoral platform and implement the general guidelines of the municipal council (\textit{Consiglio comunale}), which exercises political and administrative control of municipal activities. Like the executive committee, the municipal council is composed of a different number of elected councillors based on population size \citep{ermini2023}.\footnote{It consists of 10 councillors in municipalities with a population of fewer than 3,000 inhabitants; of 48 councillors in municipalities with more than 1 million inhabitants.}

A direct election scheme for mayors in municipalities was introduced in Italy in 1993. Before 1993, they were appointed by the municipal council; since 1993, they were directly elected under majoritarian rules by voters.\footnote{Moreover, depending on the population size of the municipalities, the most voted candidate could be elected as mayor in the first round (\emph{i.e.}, in municipalities with less than 15,000 inhabitants), or the mayor could be elected in a runoff, taking place two weeks later if none of the candidates surpasses 50\% of valid votes (\emph{i.e.}, in municipalities with more than 15,000 inhabitants).} The mayor remains in office for no more than two mandates of five years each, as stated by Law 120/1999, but the number of mandates has increased to three for municipalities with a population of up to 3,000 since 2014 (Law 56/2014).

\begin{figure}[H]
\centering 
\includegraphics[width=0.99\textwidth]{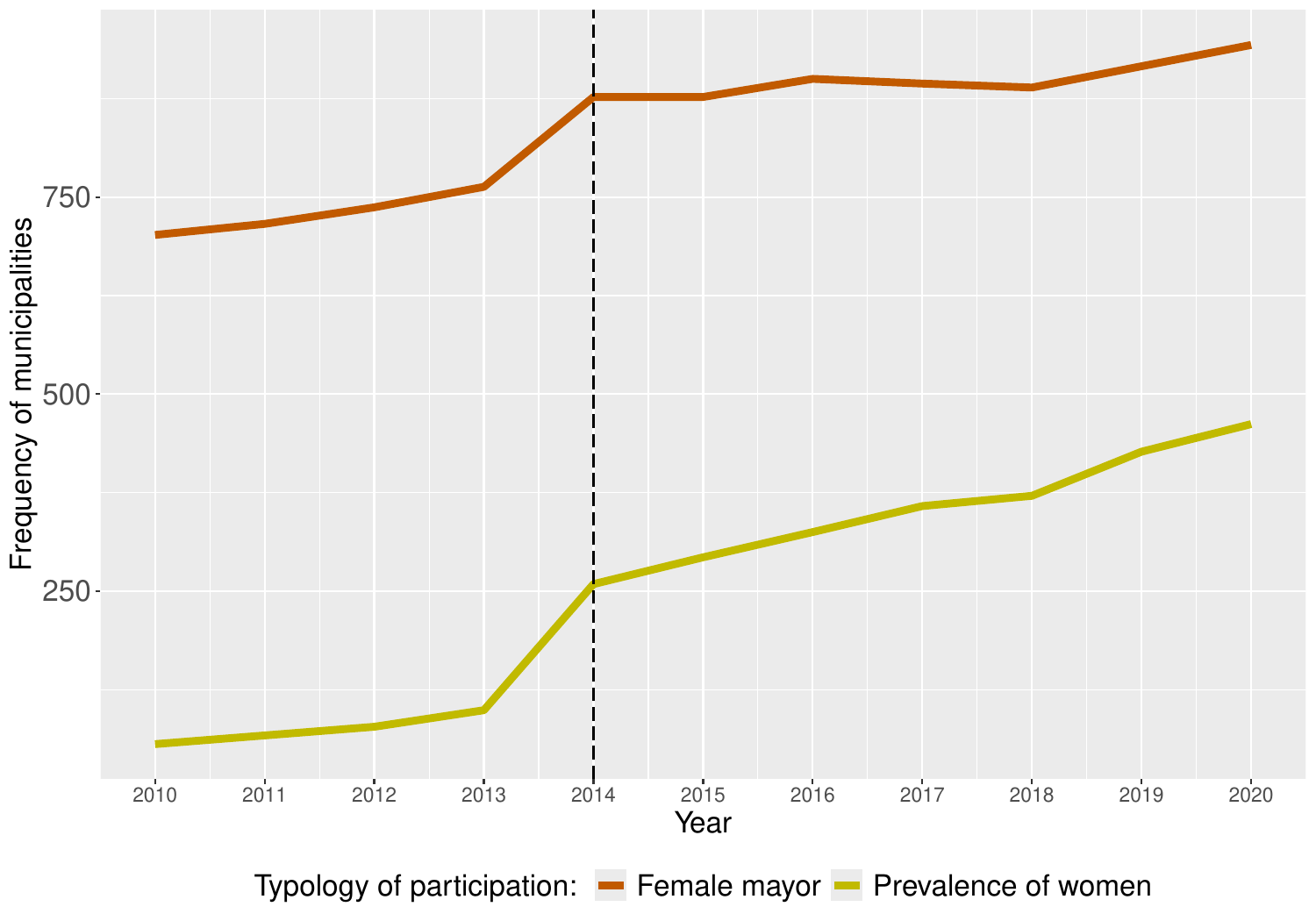}
\caption{Total number of municipalities with Female mayor and Prevalence of women by year}
\label{PGM99_plot}
\end{figure}

Recent empirical evidence from \cite{cipullo2023} shows that introducing such a direct election scheme substantially increased the proportion of municipalities where a woman was selected as the new mayor (see Figure \ref{PGM99_plot}, too). In addition, the reform increased the proportion of female elected officials, mainly in localities with a large pool of female potential candidates from which parties and voters could select.

The 1993 reform also introduced gender quotas in all municipalities having less than 15,000 inhabitants, meaning that each party was required to have at least one third of women candidates on its list of candidates for the municipal council. However, the Italian Constitutional Court abolished this part of the reform in 1995, so gender quotas were in place exclusively for three subsequent years since 1993. 

More recently, Law 215/2012 reformed the elections of Italian municipal councils by introducing a gender quota on candidates. In detail, the reform introduced both a candidate gender quota and the possibility of expressing a double preference when voting if the voters' preferences are for candidates of a different gender.\footnote{These specific features of the law were enforced exclusively in municipalities of size larger than 5,000 inhabitants.} 

Using the introduction of a gender quota by Law 215/2012 applied in 2013, \cite{andreoli2022} find that the reform affected the gender composition of candidates in Italian municipal council elections, increasing the share of female councillors of about 13.9 percentage points.\footnote{A similar finding is provided by \cite{lassebie2020} for French municipalities. A gender quota law that affects the composition of municipal councils substantially affected the share of female candidates and elected politicians but failed to promote female mayors and list leaders. Likewise, \cite{bagues2021} show that gender quotas applied to local elections in Spain boost female representation in municipal governments but do not affect mayoral positions.} Furthermore, they show that a one percentage point increase in female participation in councils raises certain expenditures (\emph{e.g.}, public order and local security) rather than others (\emph{e.g.}, administrative costs). These effects are enhanced by the increase in the share of educated, employed, and relatively young women who are municipality board members.

Finally, in 2014, the parliament approved Law 56/2014, which requires that in municipalities with more than 3,000 residents, each gender represents at least 40\% of the members of the executive committee (\textit{Giunta comunale}). Given this framework, \cite{spaziani2022} investigates the ability of the gender quota systems implemented in 1993, 2013, and 2014 to promote the election of female mayors. In general, she finds that the three policies increased the share of female politicians in less senior government positions beyond the minimum legal level. However, there is no evidence about the effects on the mayoral position, concluding that the ''acceleration effects'' \citep{obrien2016} produced by the quotas have been too weak to advance female political leadership in Italian municipal governments.






\subsection{Waste management and separate collection at the municipal level in Italy}
\label{backRifiuti}


Waste management and its separate collection have become a fundamental aspect of Italian environmental policies, which are closely related to mitigate climate change, especially in reducing greenhouse gas emissions and lowering the use of raw materials in production. If not properly managed, waste produces high air emissions. The generation of waste is strictly connected to a constant extraction of raw resources, which are necessary to produce new goods \citep{sessa2010, damato2013, musella2019} that will become new waste, so it is fundamental to design an efficient framework to dispose and manage it.
This scenario and the increased awareness of citizens and the scientific community about environmental degradation, have fostered the promotion and the adoption of more environmentally friendly disposal techniques at the European level. The environmental urgency was also recognised by the United Nations through the approval of Agenda 2030. 

Concerning Italy, although environmental and ecosystem protection was already promoted by the Constitution (Art. 117), waste disposal was effectively regulated in 1982 (with the Decree of the President of the Republic n. 915) by implementing three directives of the European Economic Community (EEC).\footnote{In detail, the directive n. 75/442 was on waste in general, the n. 76/403 was about the disposal of polychlorinated biphenyls and terphenyls, the n. 78/319 was related to toxic and harmful waste.} Overall, such a normative framework states the fundamental principles of waste management, including waste disposal, classification of waste types and allocation of responsibilities in a multilevel governance system.\footnote{The central government gives the general rules and coordinates sub-national levels; regions enact specific laws under national legislation and provide directions on separate waste collection according to local needs; provinces have lower importance, but they control the general activities, and municipalities are effectively in charge of the organisation and management of waste separate collection and its disposal, under the regulatory framework.} Under this setting, the role of municipalities is crucial for waste management \citep{agovino2016, cerqueti2021}. Local governments have the right to implement and apply the regulations enacted at the upper government level, organise waste collection services, treatment facilities, and promote awareness campaigns. 

A second important step toward reducing the effects of waste on environmental sustainability and climate change was the 1997 legislative decree (that is, the "Ronchi decree"), which reorganises the waste management legislation in Italy.\footnote{It had replaced the previous DPR and had transposed EEC Directives 91/156 on waste, 91/689 on hazardous waste, and 94/62 on packaging and packaging and packaging waste.} Two important principles on waste were introduced. First, it banned the abandonment of waste by anyone responsible for disposing of or recovering it, depending on the guidelines provided by the decree. Second, it reasserted that waste management is an activity of public interest and ruled out the recovery, reuse, and recycling of waste.\footnote{Accordingly, the quantity of waste to be reduced to disposal by explicitly calling it the "residual phase" of waste management to foster activities that envisaged a new implementation of products at the end of their life cycle (recycling, recovery to obtain raw materials from waste; waste to generate energy, etc.).} 

Finally, in 2006, the "Testo Unico Ambientale" (TUA) was introduced, representing Italy's main waste management legislation. The core of the document is the general principles (precaution, prevention, sustainability, accountability, and cooperation among those involved, according to the "polluter pays principle") and procedures of waste management (mainly the promotion, prevention, and reduction of waste production by reuse, recycling, and recovery of materials and energy from the waste itself). More important to our analysis, the TUA explicitly establishes that the municipal government is devoted to preparing a waste management plan to reduce waste, reuse it, and achieve recycling targets. 

Furthermore, the crucial role of municipalities in implementing and facilitating the recycling and recycling of materials has been reinforced by supranational directives \citep{chioatto2023}, in line with national policies to minimise the impact of climate change. 
In this framework, environmental education of citizens can be provided through information campaigns on proper waste separation and can contribute to creating a circular economy at the local level.

Ultimately, the mayor and the municipal council are responsible for defining clear targets for separate collection and recycling in the waste management plan following national and regional regulations. This implies that each municipality can decide how waste collection should be done. However, regions and metropolitan areas coordinate municipal activities by providing technical and financial support and defining guidelines and goals for separate waste collection \citep{agovino2019}. 

\section{Methods and data}
\label{method}

\subsection{Methodological approach}
\label{method2}

The empirical approach to evaluate how gender politics influences the share of collected waste at the municipal level is presented in three distinct parts, employing cutting-edge techniques to gradually dive deeper into the causal assessment of a greater female presence in the environmental behaviours and attitudes of citizens.

Within the field of causal inference and programme evaluation, the difference-in-differences (DID) approach, equivalent to two-way fixed effects regressions (TWFE) (\citealp{Chaisemartin2022}), stands as a foundational method that allows researchers to estimate the causal impact of a treatment or intervention by comparing changes in outcomes over time between a treatment group and a control group. \\
Despite its widespread use and effectiveness in many contexts, the traditional TWFE framework have some specific limitations, particularly when faced with staggered treatment timing or when treatment effects evolve gradually over time. Given this issue, many authors, including \cite{CALLAWAY2021200} and \cite{BAKER2022} propose adopting specific staggered DID models; within this staggered framework, municipalities can therefore be grouped into cohorts that initiate treatment simultaneously. \\
The approach of \cite{CALLAWAY2021200} offers several advantages over other staggered approaches to causal inference. One of these advantages is that it allows treatment effects to vary smoothly over time, accommodating both gradual and immediate changes in the impact of interventions. Furthermore, using local polynomial regression to estimate treatment effects provides more precise estimates. Furthermore, by employing a control function approach, their method effectively controls for time-varying confounders and unobserved heterogeneity. This mitigates concerns about omitted variable bias and endogeneity issues that may arise in staggered designs.

So, following the compact notation proposed by \cite{Chaisemartin2022}, we denote $\overline{Y}_{c,t}$ as the average result in period $t$ between the groups belonging to the cohort $c$, and $\overline{Y}_{n,t}$ as the average result in period $t$ between the groups that remain untreated from period 1 to $T$ (the groups never treated) for all $c$ and $t$ and for all $l \in (0,\dots,t)$.\\ 
\cite{CALLAWAY2021200} define their parameters of interest as: 
\begin{equation}
TE_{c,c+l} = E[\overline{Y}_{c,c+l}(\bm{0}_{c-1},\bm{1}_{l+1}) - \overline{Y}_{c,c+l}(\bm{0}_{c+l})]
\end{equation}
the average effect of having been treated for $l+1$ periods in the cohort that started receiving the treatment at period $c$, for every $c \in (2, ..., T)$ and $l \ge 0$ such that $l+c\le T$. \\
To estimate $TE_{c,c}$,  \cite{CALLAWAY2021200} propose\footnote{The estimators are computed by the \texttt{csdid STATA} command (\citealp{RiosAvila2021}).}:
\begin{equation}
\overline{DID}_{c,0} = \overline{Y}_{c,c} -  \overline{Y}_{c,c-1} - (\overline{Y}_{n,c} - \overline{Y}_{n,c-1})
\label{calla2}
\end{equation}
a DID estimator comparing the period $c-1$ to $c$ outcome evolution in cohort $c$ and in the never-treated groups $n$.

Up to this point, we have assumed that the outcome comparisons within cohorts and between groups are stationary in geographical space; but, in real-world scenarios, this assumption may not be fully satisfactory because the relationship between variables can vary between different geographical locations. Ignoring spatial non-stationarity can lead to biased estimates and incorrect inferences since different regions or locations may exhibit unique characteristics that affect the outcome of interest. By assuming spatial stationarity, the DID framework may fail to capture these spatial differences, potentially masking important variations in treatment effects and leading to model misspecification. \\
Incorporating rationale under Geographic Weighted Regression (GWR, \citealp{brunsdon1998geographically}) or local spatial econometric models can help address these limitations by allowing for spatially varying treatment effects and capturing spatial heterogeneity more accurately. Such models repeatedly estimate the baseline specification by varying an estimation window in which points in space are weighted according to their distance from the centre of the window. 
More specifically - at each step defined by unit $i$ - spatial weights $w_{ij}$ can be defined based on distance, such as using a linear decay function, a Gaussian kernel, or an adaptive kernel, to give more weight to nearby observations and less weight to distant ones.\\
This choice is clearly not always related to application-specific characteristics. For this reason, in Section \ref{sec_hetero} the local weights will be calculated on the basis of a restricted linear spatial decay function (see equation \ref{eq_w_linear}), while in Section \ref{sec_rob}, to empirically verify the robustness of the results obtained, the local weights will be calculated on the basis of a non-restricted Gaussian function (see Figure \ref{fig_weights}).
\begin{equation}
w_{ij} = 
\begin{cases}
 1 - (d_{ij}/d_{max}) \text{, if this ratio is $>$ 0.5} \\
 0  \text{, otherwise}
\end{cases}
\label{eq_w_linear}    
\end{equation}
where $d_{ij}$ is the Euclidean distance between the municipality $i$ and $j$ and $d_{max}$ the maximum distance between all municipalities $i$ and $j$ considered.

The staggered DID model referenced in \cite{CALLAWAY2021200} enables the calculation of ATTs for each subsequent period following the start of staggered treatments. When integrated with the previously mentioned local method, this facilitates the analysis of trends following treatment for every distinct municipality. We can then estimate clusters of curves, \emph{i.e.} similar trends between municipalities to check not only whether there is spatial stationarity on average, but also in terms of trends, \emph{i.e.} evolution over time. Functional data analysis (FDA), as defined by \cite{Ramsay2005}, is an advanced extension of traditional multivariate methodologies, specifically designed to incorporate data conceptualised as functions or curves.  Within the purview of this methodology, it is postulated that observations inhabit an infinite-dimensional continuum. However, in real-world applications, we are constrained to dealing with sampled trajectories that are accessible only at discrete temporal instances. Typically, discrete observations $X_{ij}$ of each sampled path $X_i(t)$ are obtained at a finite set of nodes $t_{ij}: j = 1 , ..., m_i$ (for example, in this application the ATT coefficients by periods after treatment; see Table \ref{tabCSDID_mayor}). Subsequently, the preliminary phase of the FDA requires the reconstitution of the data functional characteristics from these discrete observations, employing nonparametric smoothing techniques. In the context of the study described in this paper, the funFEM algorithm (\citealp{BouveyronR}) has been used to cluster different trends of the ATT post-treatment coefficient, treating them as curves within a shared and distinctive functional subspace (see \emph{e.g.} Figure \ref{PGM7_functional_cluster}).

\subsection{Data sources, main variables and descriptive statistics}
\label{sec_data}

The empirical analysis is based on annual 2010-2019 data from many sources. It relies on information related to waste services, focussing on the sorted waste collected at the municipal level, which represents the dependent variable of our analysis. Specifically, data were collected from the Italian Institute of Environmental Protection and Research (ISPRA) and include the composition of municipal waste (\emph{i.e.} sorted and unsorted) and the main cost elements for waste management functions (\emph{i.e.} related to the processing of the sorted fraction). The quantity of waste collected was calculated as the share of the total urban waste collected (sorted and unsorted). 



Concerning gender in politics, which represents the main independent variable of interest, we refer to data on municipal elections in Italy provided by the Department of Internal and Territorial Affairs of the Ministry of the Interior which annually report information on local administrators in service at the $31^{st}$ of December of each year.\footnote{It is the "Register of Local and Regional Administrators" (\textit{Anagrafe degli Amministratori Locali}), compiled yearly since 1986, which represents the individual--level register of the universe of politicians holding any office at the municipality level; \url{https://dait.interno.gov.it/elezioni/open-data}.}  The above-mentioned dataset allows us to identify the gender. Two different identifier variables (\texttt{treatment group (cohort)}) have been constructed for each municipality and year as a dichotomous variable: the presence of a female mayor and a female majority in the municipal council. 


The data set also includes geographic and socioeconomic characteristics of municipalities to account for other factors that may mediate or affect the impact of gender on the share of sorted waste collection. In detail, as highlighted by the existing literature on municipal solid urban waste management, we consider, on the one hand, geographic variables of the municipality, such as resident population (number of inhabitants), municipality area in square kilometers, population density (ratio between population and municipality area), if the municipality is a mountain municipality or not, average number of households members and the number of tourist beds \citep{StrukBod'a2022, Soukiazis2020, Romanoetal2019}. However, the estimation has taken into account the socioeconomic characteristics of the municipality as controls. Specifically, we use the treatment and recycling costs of sorted municipal waste and the collection and transport costs of sorted municipal waste \citep{Difoggiab2020}. Both variables are expressed as euros per inhabitant spent in a specific year. Finally, we have implemented the per capita average taxable incomes declared by residents of the municipality to be subject to the general income tax for individuals (IRPEF), henceforth income per inhabitant \citep{Romanoeral2022, Fiorillo2013, Romanoetal2019}. 


In summary, Table \ref{tab_summ2} reports the descriptive statistics.
\begin{table}[H] \centering 
\scalebox{0.8}{
\begin{tabular}{@{\extracolsep{5pt}}p{0.40\linewidth}rrrrr} 
\\[-1.8ex]\hline 
\hline \\[-1.8ex] 
Statistic & \multicolumn{1}{c}{N} & \multicolumn{1}{c}{Mean} & \multicolumn{1}{c}{St. Dev.} & \multicolumn{1}{c}{Min} & \multicolumn{1}{c}{Max} \\ 
\hline \\[-1.8ex] 
Sorted waste collection perc. & 62,129 & 0.522 & 0.230 & 0.000 & 1.000 \\ 
Population density & 68,608 & 276.871 & 547.262 & 0.910 & 11,494.890 \\ 
Resident population & 68,629 & 6,493.189 & 29,857.900 & 29 & 1,406,242 \\ 
Area in square kilometres & 68,620 & 34.804 & 44.229 & 0.000 & 653.822 \\ 
Average n. of household members & 68,629 & 2.295 & 0.259 & 1.100 & 3.870 \\ 
Mountain municipality (Y/N) & 68,629 & 0.075 & 0.264 & 0 & 1 \\
Number of tourist beds & 60,887 & 582.334 & 2,914.223 & 0 & 75,864 \\ 
Income per inhabitant & 68,629 & 12,344.220 & 3,052.764 & 2,688.924 & 35,452.290 \\ 
Collection/transport costs of sorted waste (euro/inhab.) & 59,756 & 35.618 & 26.312 & 1.000 & 604.200 \\ 
Treatment/recycling costs of sorted waste (euro/inhab.) & 53,523 & 13.269 & 10.743 & 1.000 & 185.450 \\ 
\hline \\[-1.8ex] 
\end{tabular} }
  \caption{Summary statistics} 
  \label{tab_summ2} 
\end{table}

\section{The empirical analysis}
\label{empirical}

We start by describing the baseline model on a global scale. This helps us to determine whether having a female mayor or a majority of female municipal councillors would affect citizens' behaviour within a staggered counterfactual framework, on average, over the whole territory. Then we investigate whether these findings are robust for different geographic areas or if omitted/hidden local factors regularly produce changing outcomes. As a final step, we check whether the effect of a majority of female municipal councillors is also stationary with respect to trends in single territories.\footnote{Findings regarding the impact of having a female mayor - being positive but less significant - are available from the authors on request.} 


\subsection{Identification strategy and  average baseline results}
\label{baseline}

As discussed in Section \ref{design}, the main objective is to pinpoint the direct influence of electing a female mayor or having a majority of women in the municipal council on improving the civic participation of citizens in environmental issues, specifically increasing the share of sorted waste collection, in line with the "civic environmentalism" approach \citep{john1994civic, agyeman2003}. 

It is well known that simple correlations might be misleading due to significant endogeneity issues, thus hindering the establishment of a reliable and robust causal link. As a baseline specification, the \texttt{percentage of sorted waste on total urban waste generated} controlling for geographic and socioeconomic variables (as described in Section 4.2) is considered for all Italian municipalities by taking into account the staggered treatment-group $TG_{it}$ and following the approach described in the equation (\ref{calla2}): 
\begin{equation}
    Perc^{sorted}_{it} = f(X^{eco}_{it}, X^{demo}_{it}, TG_{it})
\label{eq_sorted}
\end{equation}

\begin{figure}[H]
\centering 
\includegraphics[width=0.99\textwidth]{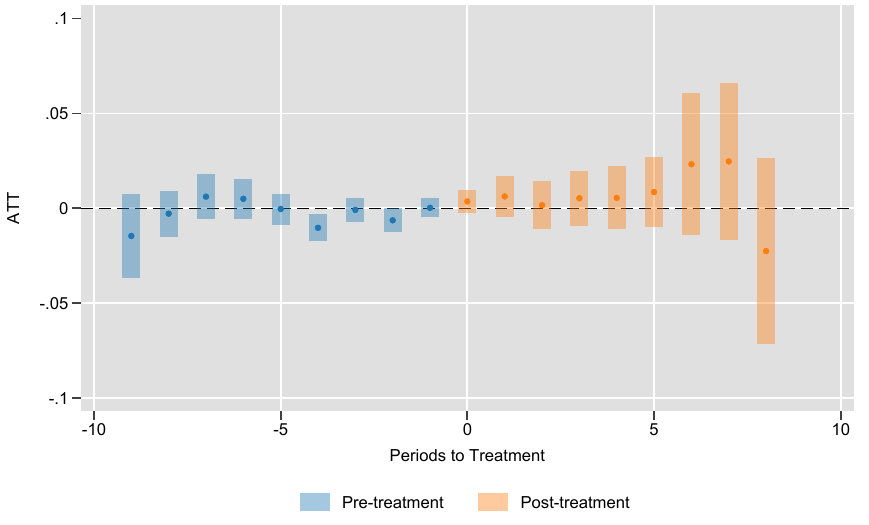}
\caption{ATT by Periods Before and After treatment, Event Study: Dynamic effects, Woman Mayor - CSDID model}
\label{completea1}
\end{figure}

The effect of electing a female mayor on changing the environmental behaviour of citizens has been first verified. Figure \ref{completea1} shows a modest positive impact, particularly in the medium term, although this impact is not statistically significant.

Table \ref{tabCSDID_mayor} also shows a positive average post-treatment coefficient (\texttt{Post\_avg}), but not significant as for post-treatment coefficients per year (\texttt{Tp0-Tp8}). The election of a female major is not sufficient to change the recycling behaviours of the municipality's citizens, even after different year from the election.
\begin{table}[H]
  \centering
\scalebox{0.8}{
    \begin{tabular}{lrrrrrr}
    \\[-1.8ex]\hline 
\hline \\[-1.8ex] 
          & \multicolumn{1}{l}{Coefficient} & \multicolumn{1}{l}{Std. err.} & \multicolumn{1}{l}{z} & \multicolumn{1}{l}{P$>$z} & \multicolumn{1}{l}{[95\% conf.} & \multicolumn{1}{l}{interval]} \\
          \\[-1.8ex]\hline 
\hline \\[-1.8ex] 
    Pre\_avg & -0.00228 & 0.001618 & -1.41 & 0.159 & -0.00545 & 0.000894 \\
    Post\_avg & 0.003923 & 0.008222 & 0.48  & 0.633 & -0.01219 & 0.020038 \\
    \midrule
    Tm9   & -0.01231 & 0.010221 & -1.2  & 0.229 & -0.03234 & 0.007726 \\
    Tm8   & -0.00154 & 0.005541 & -0.28 & 0.781 & -0.0124 & 0.009322 \\
    Tm7   & 0.002058 & 0.005473 & 0.38  & 0.707 & -0.00867 & 0.012784 \\
    Tm6   & 0.005988 & 0.004823 & 1.24  & 0.214 & -0.00346 & 0.01544 \\
    Tm5   & 0.001273 & 0.003989 & 0.32  & 0.750  & -0.00655 & 0.009092 \\
    Tm4   & -0.01235 & 0.003135 & -3.94 & 0.000     & -0.01849 & -0.0062 \\
    Tm3   & 0.00053 & 0.002917 & 0.18  & 0.856 & -0.00519 & 0.006247 \\
    Tm2   & -0.00523 & 0.002981 & -1.75 & 0.080  & -0.01107 & 0.000617 \\
    Tm1   & 0.001078 & 0.002303 & 0.47  & 0.640  & -0.00344 & 0.005592 \\
    Tp0   & 0.001847 & 0.002871 & 0.64  & 0.520  & -0.00378 & 0.007473 \\
    Tp1   & 0.003414 & 0.005087 & 0.67  & 0.502 & -0.00656 & 0.013384 \\
    Tp2   & -0.00247 & 0.006115 & -0.4  & 0.686 & -0.01446 & 0.009512 \\
    Tp3   & 0.001258 & 0.007017 & 0.18  & 0.858 & -0.0125 & 0.01501 \\
    Tp4   & 0.003119 & 0.00805 & 0.39  & 0.698 & -0.01266 & 0.018897 \\
    Tp5   & -0.00042 & 0.008901 & -0.05 & 0.962 & -0.01787 & 0.017022 \\
    Tp6   & 0.020741 & 0.018081 & 1.15  & 0.251 & -0.0147 & 0.056179 \\
    Tp7   & 0.024561 & 0.020136 & 1.22  & 0.223 & -0.0149 & 0.064026 \\
    Tp8   & -0.01674 & 0.024717 & -0.68 & 0.498 & -0.06518 & 0.031707 \\
    \\[-1.8ex]\hline 
\hline \\[-1.8ex] 
    \end{tabular}}
      \caption{ATT by Periods Before and After treatment, Event Study: Dynamic effects, Woman Mayor - CSDID model}
  \label{tabCSDID_mayor}%
\end{table}%

\begin{figure}[H]
\centering 
\includegraphics[width=0.99\textwidth]{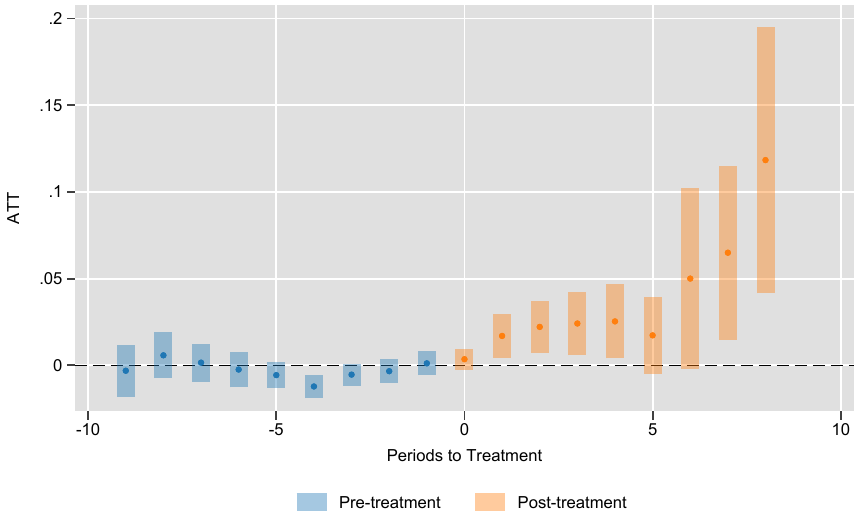}
\caption{ATT by Periods Before and After treatment, Event Study: Dynamic effects, Majority of female councillors CSDID model}
\label{completeb1}
\end{figure}

Shifting focus from the election of a woman mayor to a more evident change, like the election of a predominantly female council, the argument appears to be much more persuasive. In contrast to a very flat pretrend, Figure \ref{completeb1} demonstrates a distinct positive and statistically significant impact over nearly all years on the share of sorted waste collection; this effect becomes more pronounced over time.

Table \ref{tabCSDID_preval} shows a statistically significant average effect of around 3\%, which is also corroborated by the average post-treatment coefficients, which are positive and almost all statistically significant. 
\begin{table}[H]
  \centering
\scalebox{0.8}{
    \begin{tabular}{lrrrrrr}
    \\[-1.8ex]\hline 
\hline \\[-1.8ex] 
          & \multicolumn{1}{l}{Coefficient} & \multicolumn{1}{l}{Std. err.} & \multicolumn{1}{l}{z} & \multicolumn{1}{l}{P$>$z} & \multicolumn{1}{l}{[95\% conf.} & \multicolumn{1}{l}{interval]} \\
    \\[-1.8ex]\hline 
\hline \\[-1.8ex] 
          
    Pre\_avg & -0.00262 & 0.001555 & -1.69 & 0.092 & -0.00567 & 0.000424 \\
    Post\_avg & 0.038078 & 0.009995 & 3.81  & 0.000     & 0.018489 & 0.057668 \\
    \midrule
    Tm9   & -0.00312 & 0.007667 & -0.41 & 0.684 & -0.01814 & 0.011909 \\
    Tm8   & 0.005773 & 0.006723 & 0.86  & 0.391 & -0.0074 & 0.018949 \\
    Tm7   & 0.001562 & 0.005591 & 0.28  & 0.780  & -0.0094 & 0.012521 \\
    Tm6   & -0.00242 & 0.005265 & -0.46 & 0.645 & -0.01274 & 0.007895 \\
    Tm5   & -0.00564 & 0.003758 & -1.5  & 0.134 & -0.013 & 0.001727 \\
    Tm4   & -0.01219 & 0.003449 & -3.53 & 0.000     & -0.01895 & -0.00543 \\
    Tm3   & -0.00536 & 0.003213 & -1.67 & 0.095 & -0.01165 & 0.00094 \\
    Tm2   & -0.00338 & 0.003471 & -0.97 & 0.331 & -0.01018 & 0.003427 \\
    Tm1   & 0.001163 & 0.003489 & 0.33  & 0.739 & -0.00568 & 0.008 \\
    Tp0   & 0.003533 & 0.00305 & 1.16  & 0.247 & -0.00244 & 0.00951 \\
    Tp1   & 0.016952 & 0.006371 & 2.66  & 0.008 & 0.004464 & 0.029439 \\
    Tp2   & 0.022156 & 0.007557 & 2.93  & 0.003 & 0.007344 & 0.036967 \\
    Tp3   & 0.024123 & 0.009401 & 2.57  & 0.010  & 0.005698 & 0.042548 \\
    Tp4   & 0.02533 & 0.010912 & 2.32  & 0.020  & 0.003944 & 0.046716 \\
    Tp5   & 0.017294 & 0.01134 & 1.52  & 0.127 & -0.00493 & 0.03952 \\
    Tp6   & 0.050024 & 0.026545 & 1.88  & 0.059 & -0.002 & 0.10205 \\
    Tp7   & 0.064944 & 0.025542 & 2.54  & 0.011 & 0.014883 & 0.115005 \\
    Tp8   & 0.118349 & 0.039199 & 3.02  & 0.003 & 0.041519 & 0.195178 \\

        \\[-1.8ex]\hline 
\hline \\[-1.8ex] 
    \end{tabular}}
      \caption{ATT by Periods Before and After treatment, Event Study: Dynamic effects, Majority of female councillors CSDID model}
  \label{tabCSDID_preval}%
\end{table}%

Analysing an econometric model from the particular to the general ensures a precise understanding of each variable before exploring their broader interactions and impacts. For this reason, different specifications of the model, which are described in Table \ref{tabmodelli} of \ref{app1}, have been estimated. Figures \ref{plotWomanMajor} and \ref{plotPreWoman} graphically report the obtained results.


So far, we have supported recent studies that emphasise the crucial role of women as "green" policy makers and advocates for public investments in climate change mitigation and adaptation policies \citep{ramstetter2019, mavisakalyan2019, hessami2020, casarico2022}, specifically when the municipal council is predominantly made up of females. 

In the following section, by focusing on female municipal council only, we will go a step further by examining the effect of social and cultural geography on these general results.

\subsection{Spatial non-stationarity}
\label{sec_hetero}

Standard counterfactual approaches assume that local factors are captured by variables specified in the model or by individual spatial dummies, thereby assuming that unexplained heterogeneity is saturated.

The proposed approach recognises that hidden local factors might influence the outcome of policies. Hence, instead of imposing specific ex ante geography or proxies for territorial control variables, we move the estimation window continuously over the territory, checking whether the estimated coefficients remain stable in all spatial windows. If they do not remain stable, the policy has a locally differential effect that is linked to specific spatial factors. 

This approach mitigates the need to verify a fundamental hypothesis in counterfactual models, specifically the "\emph{no hidden confounders}" assumption, which requires all significant confounding variables to be identified and incorporated into the analysis. 

\begin{figure}[H]
\centering 
\includegraphics[width=0.99\textwidth]{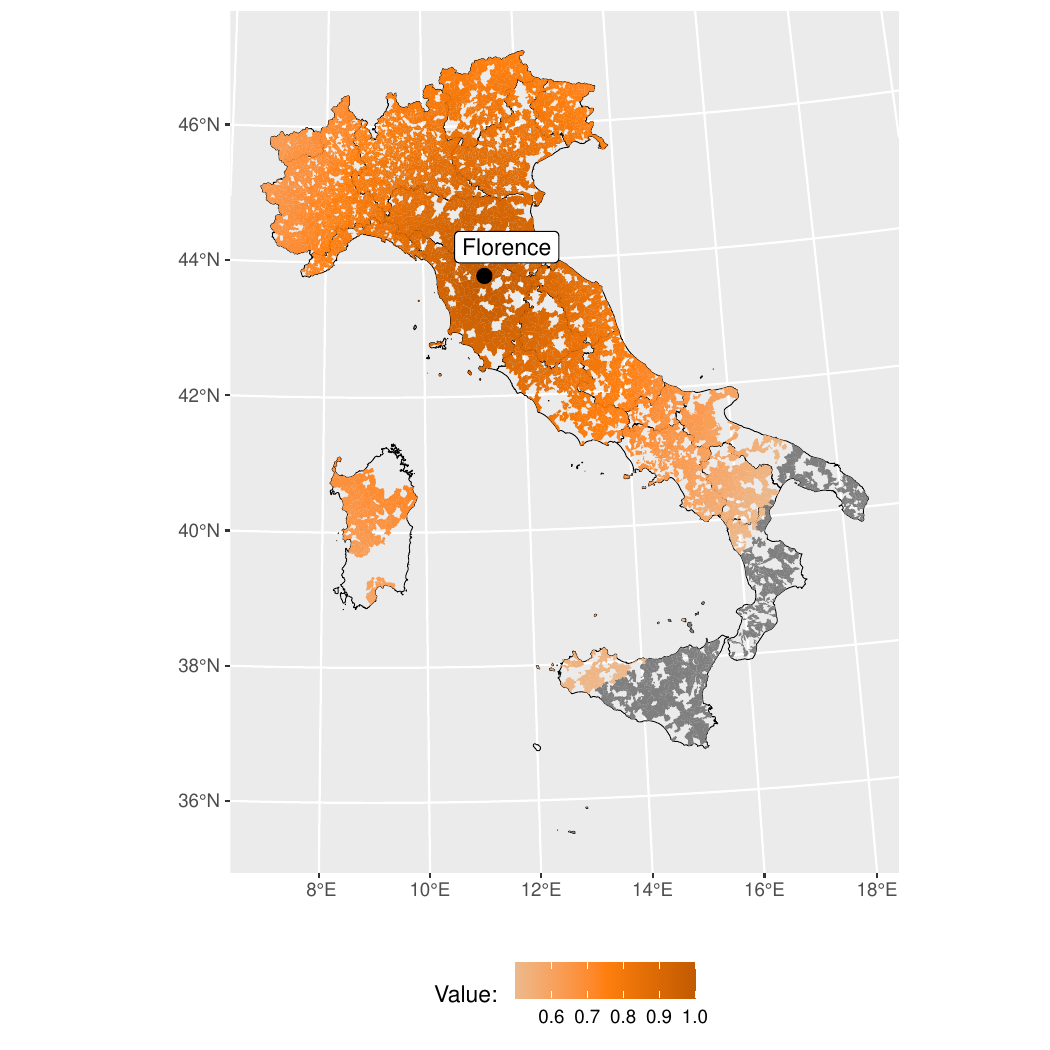}
\caption{Local linear weights, Municipality = Florence}
\label{PGM5b_pesiGWR}
\end{figure}

Given these premises, the GWR-like linear weights have been calculated for each municipality $i$ compared to the other 6,239 municipalities $j$ at each step $i$. Figure \ref{PGM5b_pesiGWR} shows, for example, the weights centred in Florence; in this case, the weight is equal to 1 for Florence and decreases as the distance of each municipality $j$ from Florence increases.

By positioning the spatial window in the centre of each municipality and applying the corresponding weights at each stage, we no longer calculated a single average post-treatment coefficient, but 6,239 average post-treatment coefficients to assess the spatial stationarity of the global model (see Table \ref{tabCSDID_preval} and Figure \ref{completeb1}). 

\begin{figure}[H]
\centering 
\includegraphics[width=0.99\textwidth]{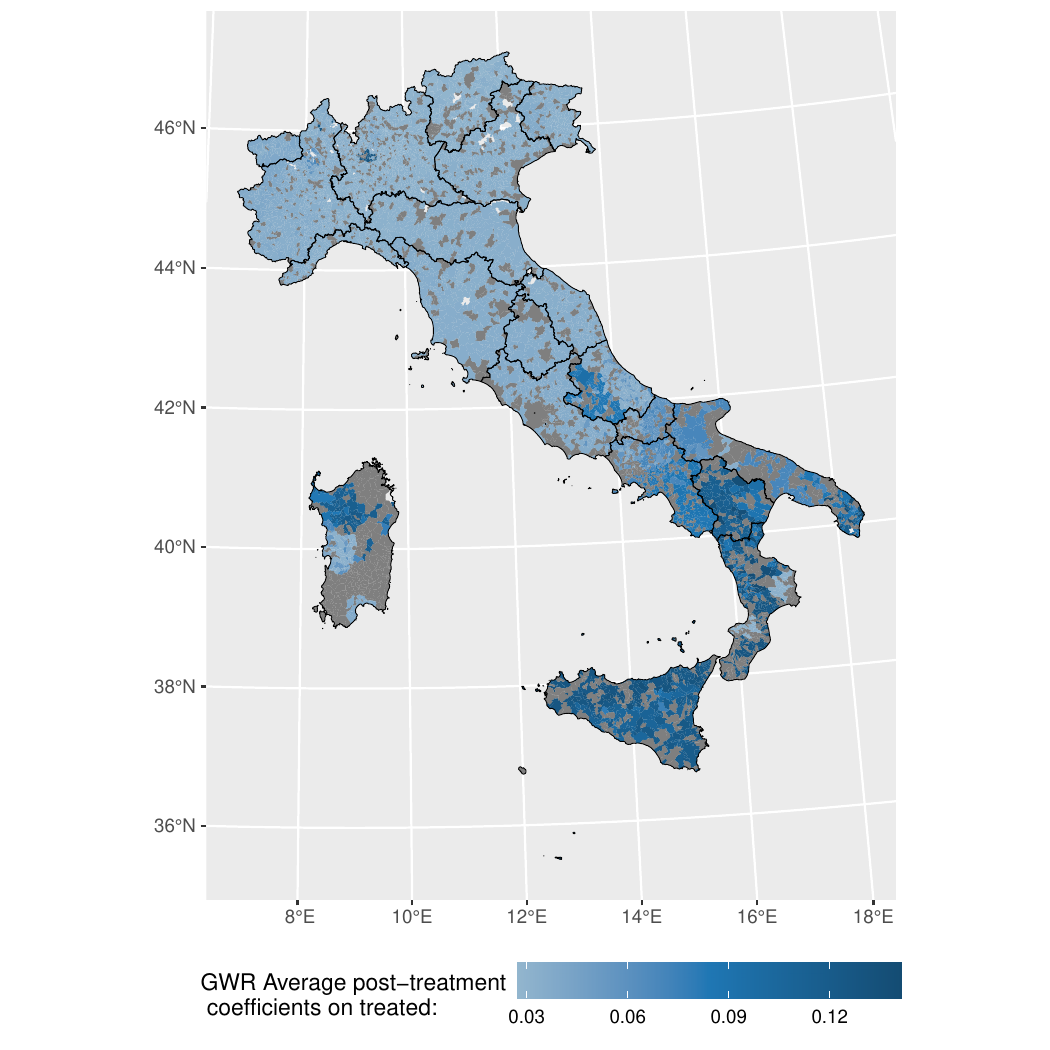}
\caption{Map of the GWR ATT post-treatment coefficients, Majority of female councillors}
\label{PGM7_sfplot1}
\end{figure}

Based on Figure \ref{PGM7_sfplot1}, we face a two-fold effect depending on where it occurs in the territory. The first underlying effect suggests that greater female representation leads to greater attention to sustainability-related issues; the magnitude corresponds to an increase of around 3\% and it is very similar to the result reported in the global model (see Table \ref{tabCSDID_preval}). This effect is very homogeneous in the Centre-North municipalities. The second additional effect, which is very strong and localised, reflects the positive impact of the majority of female councillors experienced in the South of Italy and Islands; the change most favourably impacted these two areas.


To verify whether these highly heterogeneous impacts are substantial or simply due to anomalous and transient factors, it becomes essential to investigate the presence of clear spatial differences in trends. Using the clustering procedure for individual trends, here considered as individual functional curves, proposed by \cite{BouveyronR} and described in Section \ref{method2}, we can identify three distinct clusters of curves depicting the heterogeneous effect of the increased presence of women in municipal councils on the sorted waste collection percentage of municipality citizens over years.
\begin{figure}[H]
\centering 
\includegraphics[width=0.99\textwidth]{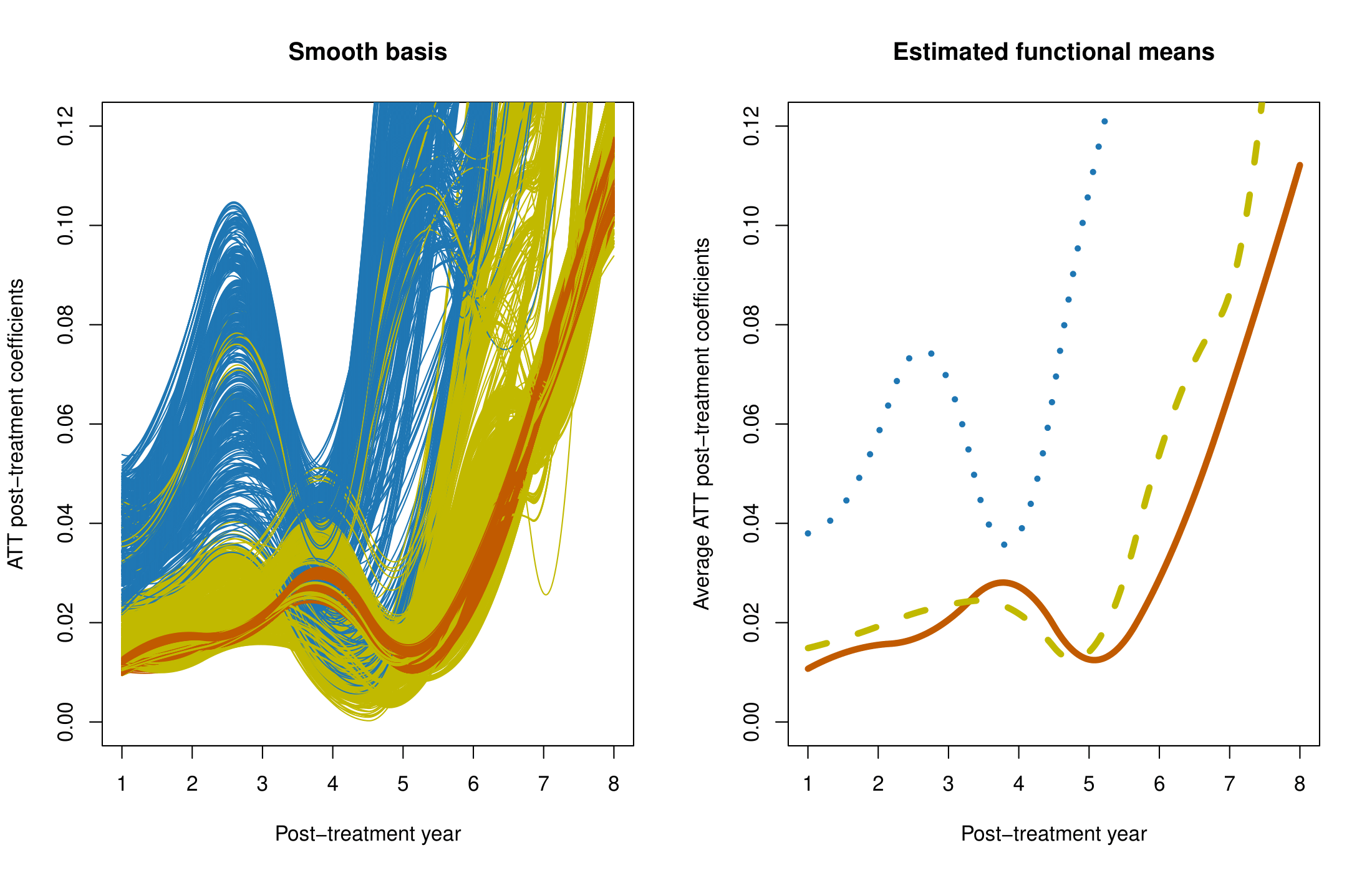}
\caption{Smoothed trend basis and functional clusters of ATT post-treatment effects, Majority of female councillors}
\label{PGM7_functional_cluster}
\end{figure}

\begin{figure}[H]
    \centering
    \subfloat[\centering Map]{{\includegraphics[width=6cm]{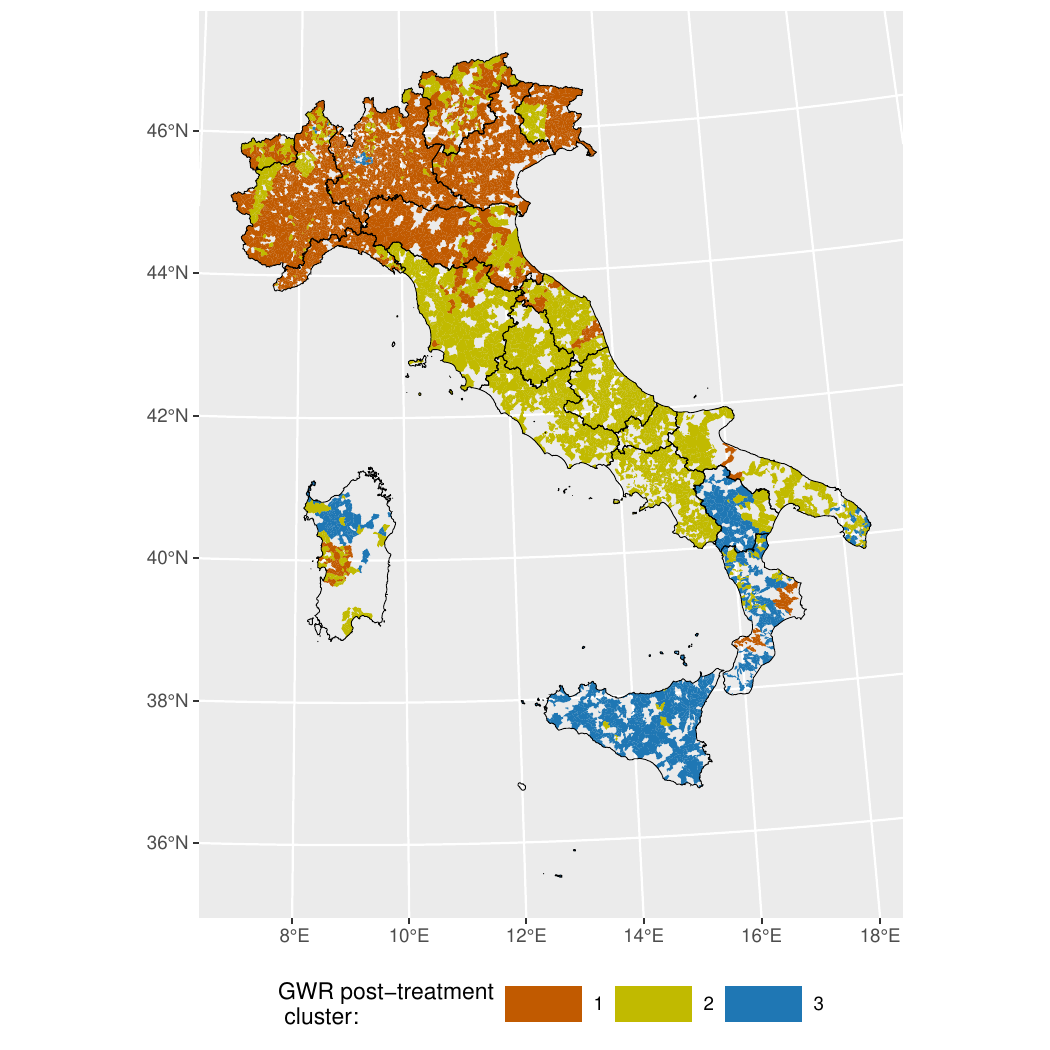} }}%
    \qquad
    \subfloat[\centering Distribution]{{\includegraphics[width=6cm]{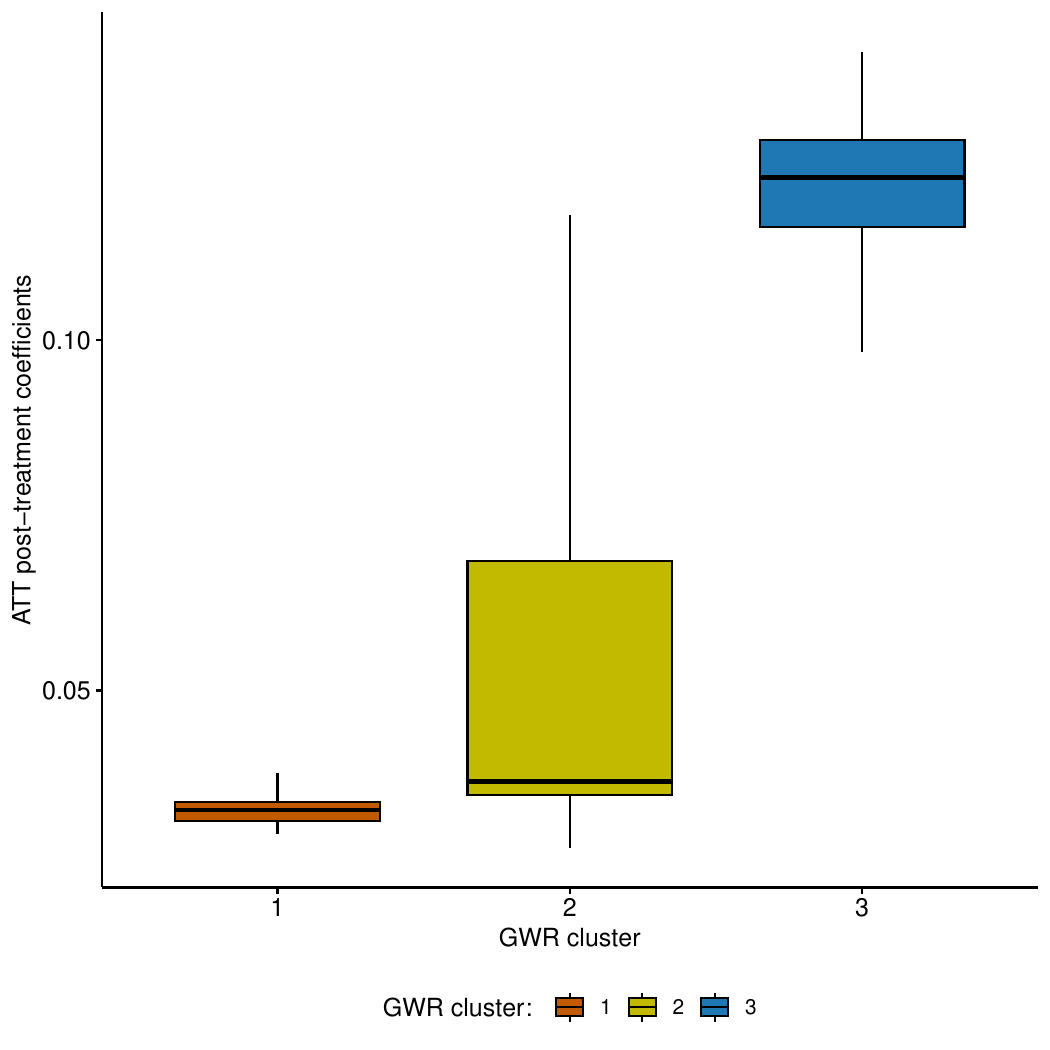} }}%
    \caption{Map and distributions of the ATT post-treatment coefficients by functional cluster, Majority of female councillors }%
    \label{fig_clu}%
\end{figure}

Figure \ref{PGM7_functional_cluster} shows the smoothed trend basis (a curve for each municipality) on the left and the estimated functional clusters of the trends of the effect of ATT post-treatment on the right. Three patterns emerge: \emph{i)} Cluster 1 (referred to Figure \ref{fig_clu}), mainly involving northern municipalities, is identified as the most uniform, with a minor but notable influence on the prevalence of women in the municipal council. Cluster 2, encompassing the Centre, Campania, and Apulia, is the least uniform, exhibiting a moderate effect, though not markedly trend divergent from Cluster 1. Finally, Cluster 3, which mainly comprises Calabria and Sicily, shows a more pronounced effect. \emph{ii)} The impact across the three regions appears to be transient, i.e. more pronounced initially and diminishing in years 4 and 5; \emph{iii)} a significant increase in the effect is observed toward the last years of the analysis (years 7 and 8), but this should be moderated considering the expanding confidence intervals during those years; additionally, in the early years of the panel, there were very few municipalities with a prevalence of women in the municipal council (see Figure \ref{PGM99_plot}). 


When estimating the impact of female mayors alone, the GWR results are weakly positive but not statistically significant in all areas of Italy. Therefore, this stationary trend does not allow clear differences to emerge.\footnote{Results are not reported and available upon request from the authors.}
\begin{figure}[H]
\centering 
\includegraphics[width=0.99\textwidth]{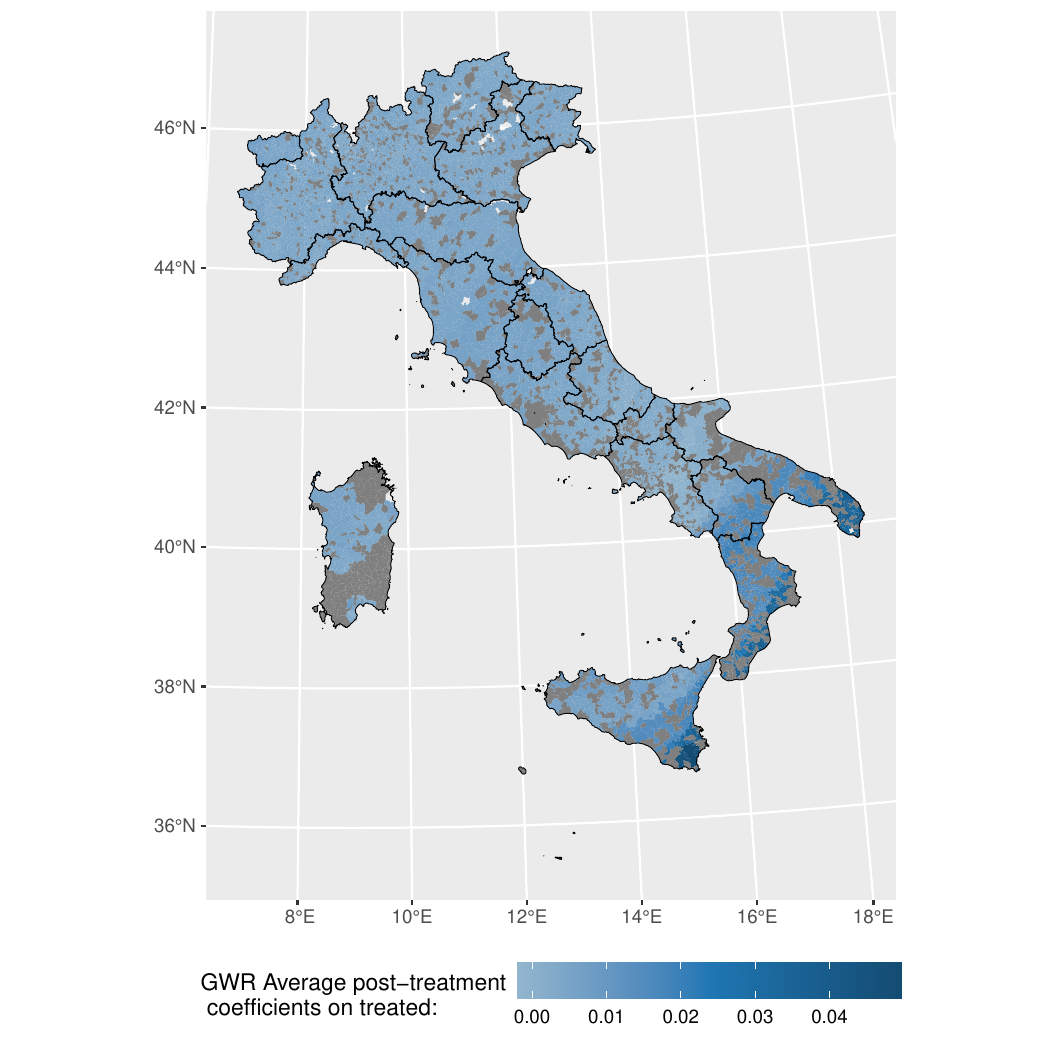}
\caption{Map of the GWR ATT post-treatment coefficients, Woman Mayor}
\label{PGM8_sfplot1}
\end{figure}

\subsection{Robustness analysis}
\label{sec_rob}

Robustness checks are implemented by modifying the local weights through a Gaussian decaying nonlinear weight function as described in Equation (\ref{eq_w_GAUSS}):
\begin{equation}
    w_{ij} = exp(-(d_{ij}/h)^2)
\label{eq_w_GAUSS}    
\end{equation}

where the distance between the municipality $i$ and $j$ ($d_{ij}$) is standardised for the bandwidth $h$ calculated for a given geographically weighted regression by optimising the linear model introduced in equation (\ref{eq_sorted}).

This weighting approach, unrestricted and featuring broader tails (see Figure \ref{fig_weights}.b), incorporates more units at each stage than the earlier scheme. Consequently, we expect that the disparities between regions will be smoother.

In Figure \ref{PGM11_sfplot1} in \ref{app3}, we observe a trend similar to the estimation using linear weights, where the prevalence of the effect in southern regions is highlighted. By construction, this evidence is not as pronounced as in Figure \ref{PGM7_sfplot1}. However, the correlation between the post-average coefficients computed across various municipalities in the constrained linear scenario and the unconstrained Gaussian scenario is impressively high at 0.711. This substantial correlation underlines the robustness of our estimates, indicating a reliable inference despite differing methodologies.

\subsection{Implications of spatial non-stationarity}
\label{sec_second}


The non-stationarity of the estimated impacts highlights that changes in societal behaviour and governance, including environmental ones, are dynamic and require enduring cultural efforts to achieve lasting transformation. Those changes are mainly influenced by rooted social and cultural aspects \citep{Massey1994, guiso_etal_2006, guiso_etal_2011} rather than administrative boundaries. 

One of the sociocultural features in Italy is the strong role of the family,
 fostering a unique form of social capital where trust and mutual support are paramount and whose influence extends to various aspects of life, affecting social behaviours, economic decisions, and political attitudes \citep{mareetal2024}. Within the Italian family, women play a crucial role \citep{caldwell1991italian, willson2009women}; initially, they were the ones who performed the so-called ''traditional'' tasks (e.g., care of children and home), more recently, since the postwar period, women have started to be involved in more complex tasks or activities carried out by men until that moment (e.g., working). Changes in the social role of women have been heterogeneous in different regions, reflecting diverse cultural and economic contexts.

All this given, the role of women in family and society represents a case of non-stationarity in sociocultural geography, pointing out that the impact of changes on an important environmental issue is not uniform but varies in strength and duration. The transition has been smooth in some regions and has been integrated into the existing social background. In others, where traditional views on gender roles are more rooted, changes have been more disruptive.

Our research highlights two main findings on the changing roles of women in municipal governance in Italy. Municipalities have responded positively to the changing roles of women by diversifying their compositions. This is more pronounced in areas where traditional gender roles were previously more rigid. The disruptive nature of these changes in these areas catalyses a more visible transformation in local governance structures.

Despite the initial positive response, the impact of the change tends to fade in the medium term (after 5-6 years). This suggests that while initial shifts in governance and social attitudes occur, also following regulatory measures \citep{depaola2014}, it is useless if they are not supported by continuous cultural efforts. In other words, long-term values, beliefs, and traditions play a significant role in this process. They might counterbalance in the medium-long term, indicating the need for cultural actions to maintain and build upon these positive changes.

\section{Concluding remarks}
\label{final}

In the last decades, the role of women in society, the growing environmental focus, and their interaction have received increasing attention. Changes related to these aspects are strictly connected to local territories and their institutions, being able to affect culture and citizens' habits through public policies.




From a policy perspective, our results underscore the critical role of gender in shaping citizens' environmental attitudes and behaviours. Bridging the gender gap in political representation can significantly contribute to achieving the green transition goals. In addition, understanding the geographical nuances of these impacts is crucial for designing effective and context-sensitive policies. Our study highlights the intersection of gender politics and environmental sustainability, advocating for greater female representation in political spheres as a catalyst for positive environmental change. 

Following this direction, local authorities could foster, on the one hand, diversity and inclusion initiatives. On the other hand, citizens and municipalities are the key levers of change; local governments are the most responsible entities for making local areas more dynamic and institutions more gender-equal, inclusive, and democratic rather than more environmentally sustainable. Put differently, considering gender issues is crucial for the success and efficiency of climate change and environmental policies. Ignoring these aspects can make climate actions less effective, as they may only engage a portion of the population or potentially harm specific social groups.

Given the supportive results of women's political roles in higher ecological sustainability, how female politicians can manage and address ongoing global issues remains to be investigated. Future research should continue to explore the long-term effects of such representations and consider other factors that may influence these dynamics.


\small
\bibliography{paper}
\bibliographystyle{elsarticle-harv}


\appendix
\setcounter{table}{0}
\setcounter{figure}{0}

\section{Spatial decay functions}
\label{app0}

\begin{figure}[H]
    \centering
    \subfloat[\centering Linear]{{\includegraphics[width=9cm]{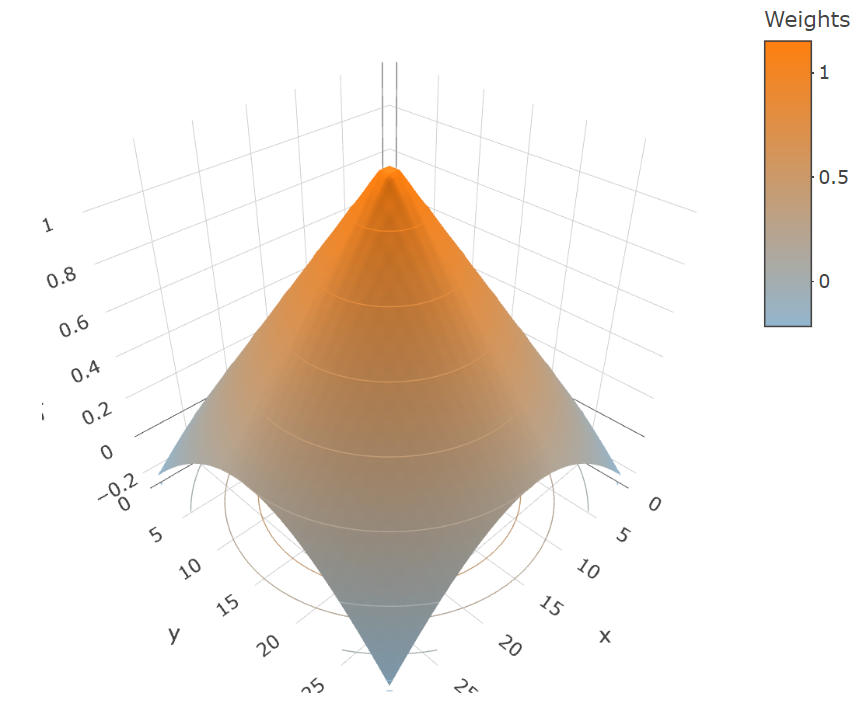} }}%
    \qquad
    \subfloat[\centering Gaussian]{{\includegraphics[width=9cm]{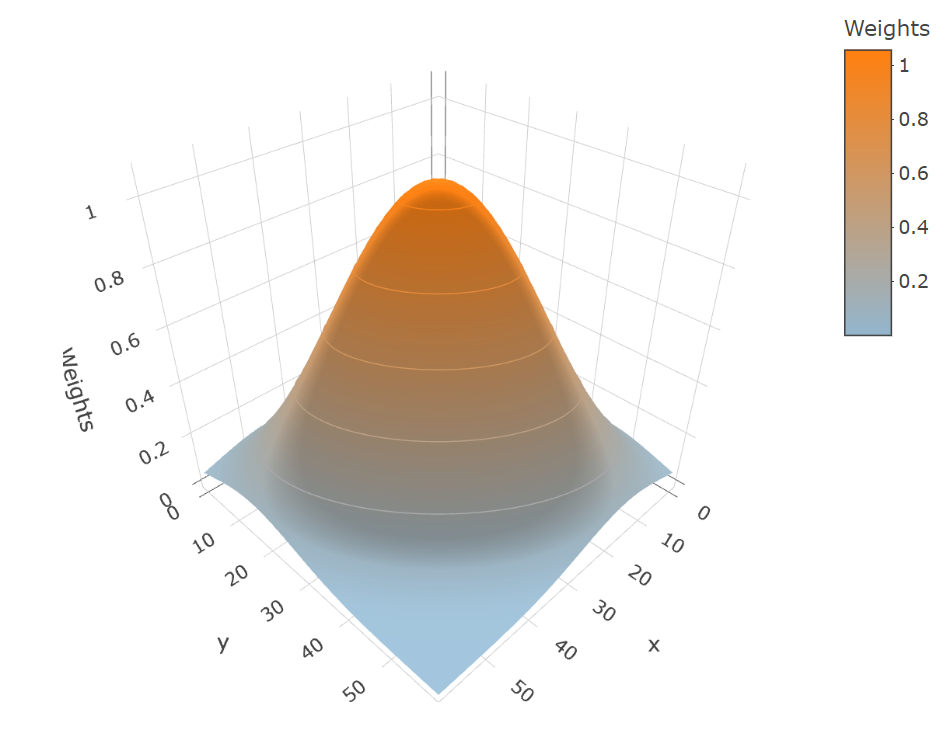} }}%
    \caption{Linear and Gaussian weights}%
    \label{fig_weights}%
\end{figure}

\section{From baseline to complex model}
\label{app1}

In the appendix, the Average Treatment Effects (ATT) for periods before and after treatment are separately documented for scenarios with a Woman Mayor and where there is a majority of female councillors (different $TG$), under different specifications. \medskip

\begin{table}[htbp]
  \centering
  \scalebox{0.8}{
    \begin{tabular}{p{0.58\linewidth}p{0.58\linewidth}}
           \\[-1.8ex]\hline 
\hline \\[-1.8ex] 

    \multirow{2}[0]{*}{Model} & Specification \\
          & $Perc^{sorted} = f(TG \text{ and covariates})$ \\
    
           \\[-1.8ex]\hline 
\hline \\[-1.8ex]

    Baseline  & No covariates \\
    \midrule
    Baseline and Socio-Demographic Controls 1  & Density +  Mountain +  Tourist beds   \\
      \midrule
    Baseline and Socio-Demographic Controls 2  & Density +  Area +  Number of household members +  Mountain +  Tourist beds \\
      \midrule
    Baseline and Socio-Demographic Controls 3  & Density +  Population +  Area +  Number of household members +  Mountain +  Tourist beds \\
      \midrule
    Baseline and Economic Controls 1  & Collection/transport cost per hab. +  Treatment/recycling costs per hab.   \\
      \midrule
    Baseline and Economic Controls 2  & Collection/transport cost per kg +  Treatment/recycling costs per kg   \\
      \midrule
    Baseline and Economic Controls 3  & Income per capita   \\
      \midrule
    Baseline and Economic Controls 4  & Income per capita +  Collection/transport waste per hab. +  Treatment/recycling costs per hab.  \\
      \midrule
    Baseline and Economic Controls 5  & Income per capita +  Collection/transport waste cost per kg +  Treatment/recycling costs per kg   \\
      \midrule
    Complete  & Density +  Population +  Area +  Number of household members +  Mountain +  Tourist beds Income per capita +  Collection/transport cost per hab. +  Treatment/recycling costs per hab.  \\

            \\[-1.8ex]\hline 
\hline \\[-1.8ex]
    \end{tabular}}
    \caption{From baseline to complex counterfactual model}
  \label{tabmodelli}%
\end{table}%

\begin{figure}[H]
\centering 
\includegraphics[width=0.99\textwidth]{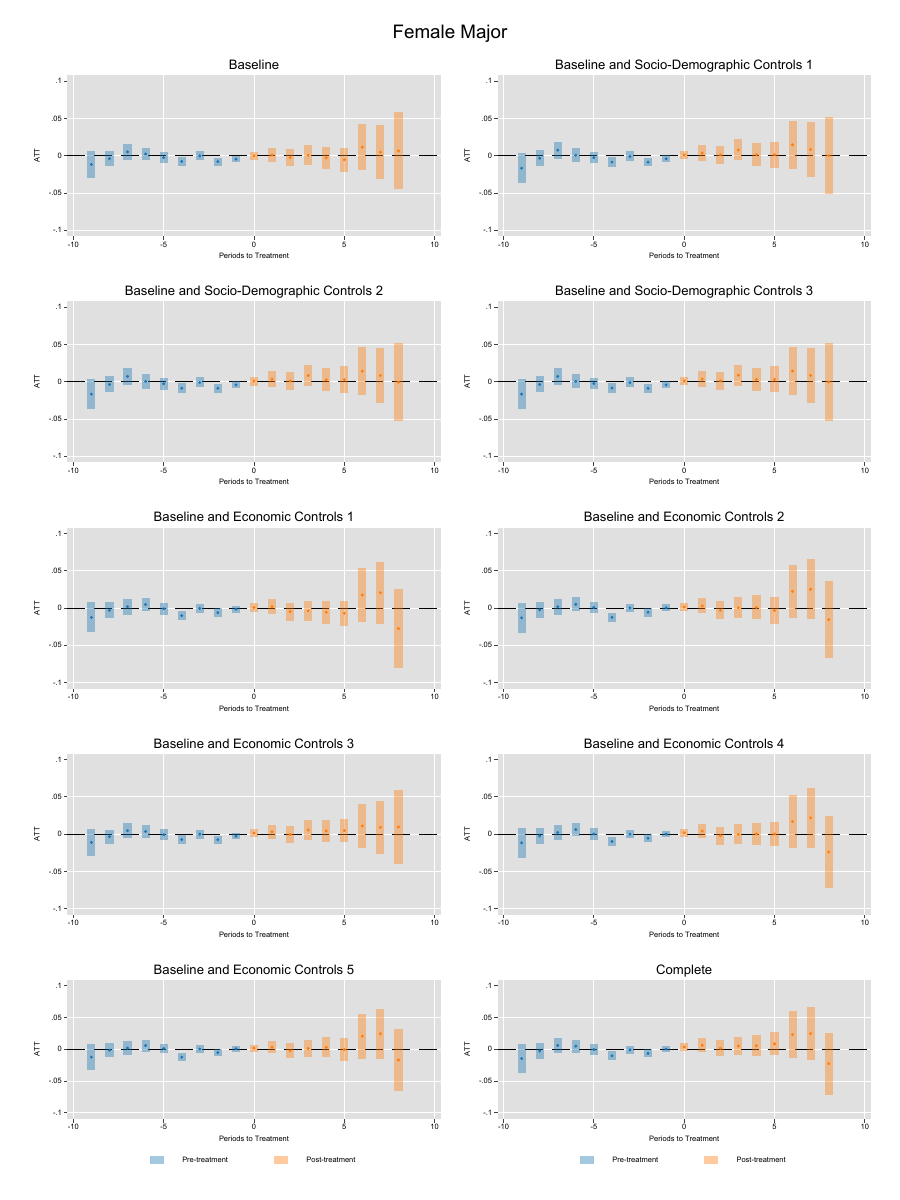}
\caption{ATT by Periods Before and After treatment, Event Study: Dynamic effects, Woman Mayor - CSDID model, by specification}
\label{plotWomanMajor}
\end{figure}

\begin{figure}[H]
\centering 
\includegraphics[width=0.99\textwidth]{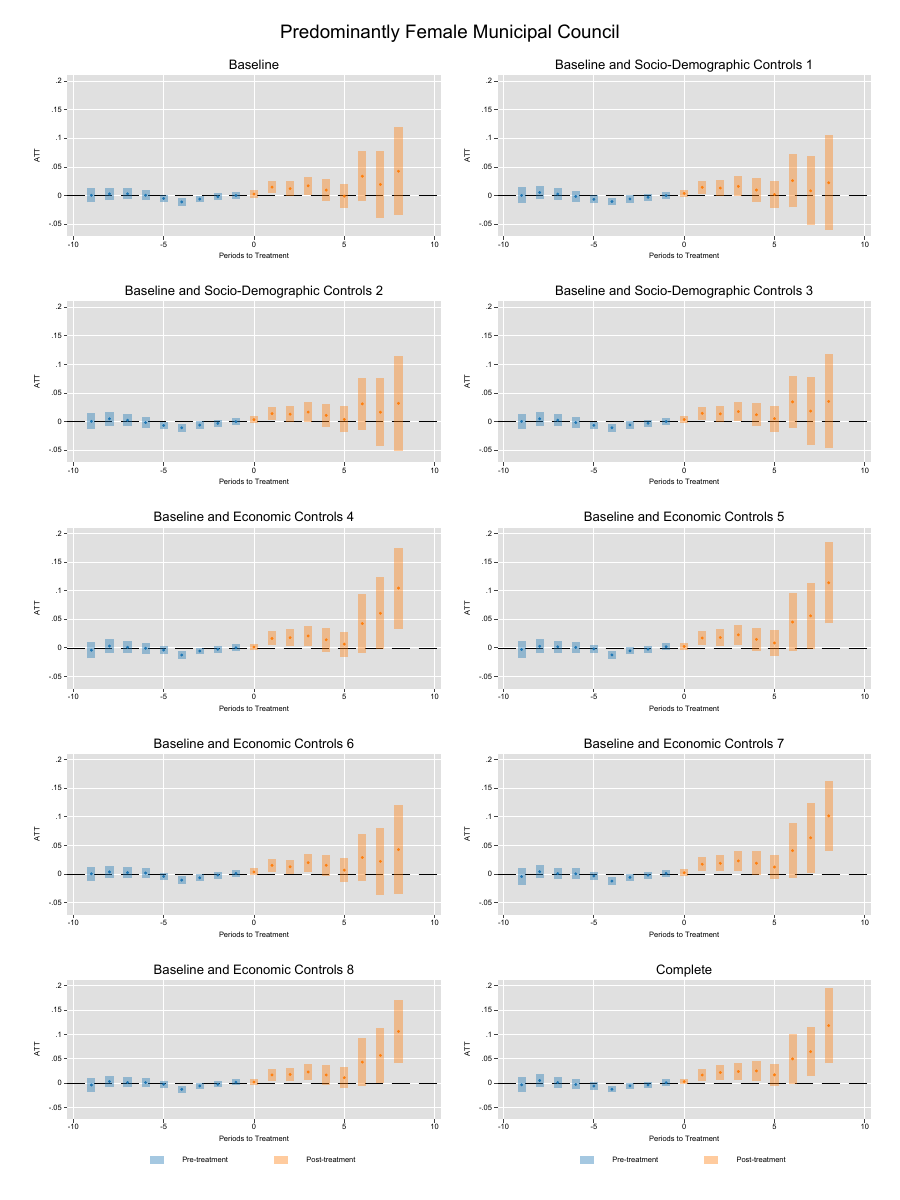}
\caption{ATT by Periods Before and After treatment, Event Study: Dynamic effects, Majority of female councillors CSDID model, by specification }
\label{plotPreWoman}
\end{figure}

\section{Robustness}
\label{app3}

\begin{figure}[H]
\centering 
\includegraphics[width=0.99\textwidth]{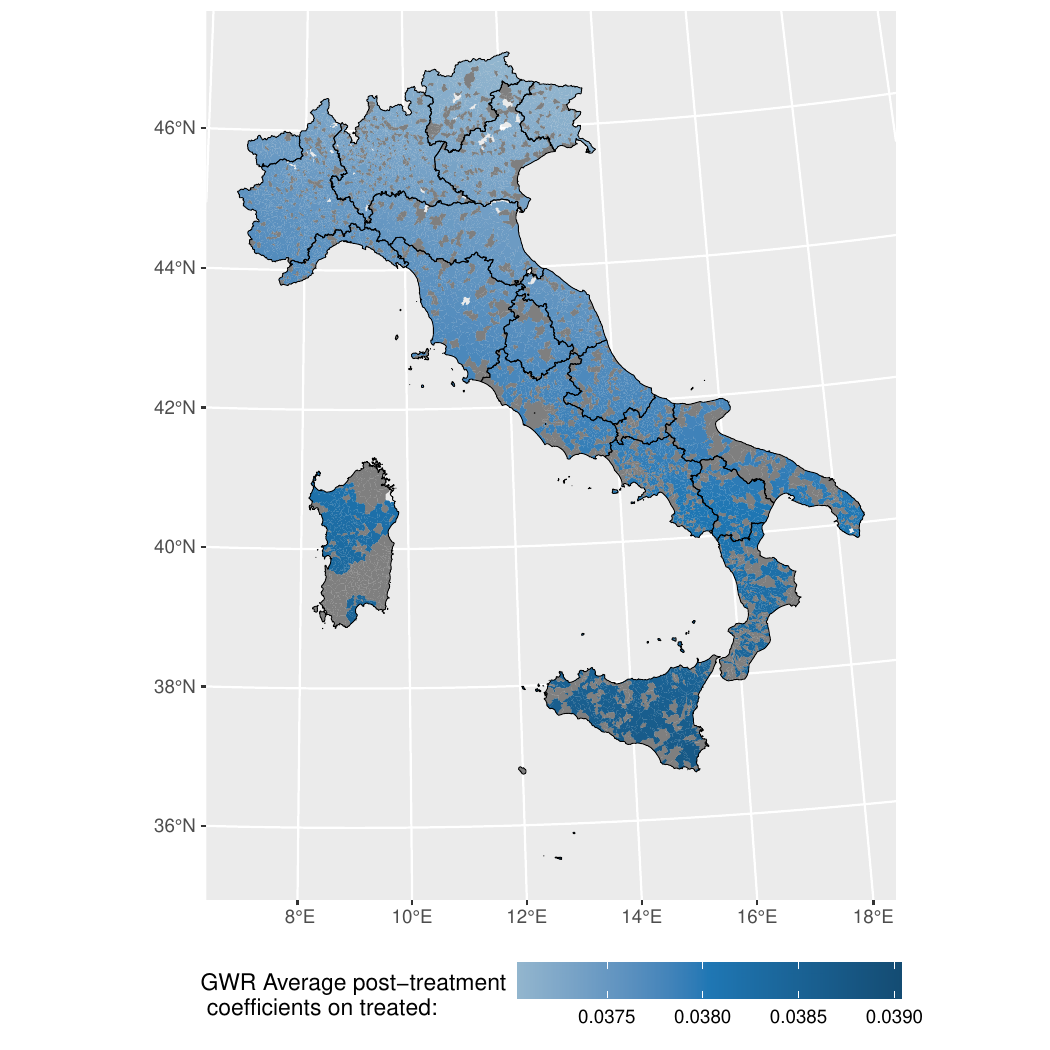}
\caption{Map of the GWR ATT post-treatment coefficients, Woman Mayor, Gaussian weights}
\label{PGM11_sfplot1}
\end{figure}

\end{document}